\documentclass[reqno]{amsart}

\usepackage[utf8]{inputenc}
\usepackage{float}
{ % táblázat vertikális nyújtása
\usepackage{makecell}
\usepackage{graphicx}
{ % táblázat vertikális nyújtása
\usepackage{color}

\begin{document}

\title{On the rarefied gas experiments}
  \author{Róbert Kovács$^{1,2,3}$}
  
\address{
$^{1}$  Department of Energy Engineering, Faculty of Mechanical Engineering, BME, Budapest, Hungary\\
$^{2}$   Department of Theoretical Physics, Wigner Research Centre for Physics,
Institute for Particle and Nuclear Physics, Budapest, Hungary \\
$^3$  Montavid Thermodynamic Research Group
}

\begin{abstract}
There are limits of validity of classical constitutive laws such as Fourier and Navier-Stokes equations, phenomena beyond those limits have been experimentally found many decades ago. However, it is still not clear what theory would be appropriate to model different non-classical phenomena under different conditions considering either the low-temperature or composite material structure. In this paper, a modeling problem of rarefied gases is addressed. It covers the mass density dependence of material parameters, the scaling properties of different theories and aspects of how to model an experiment. In the following, two frameworks and their properties are discussed. One of them is the kinetic theory based Rational Extended Thermodynamics; the other one is the non-equilibrium thermodynamics with internal variables and current multipliers. In order to compare these theories, an experiment performed by Rhodes is analyzed in detail. It is shown that the density dependence of material parameters has a severe impact on modeling capabilities and can lead to very different results.
\end{abstract}

\maketitle
\pagestyle{plain}
\section{Introduction} 

The classical material laws such as Fourier and Navier-Stokes are acceptable for tasks concerning homogeneous materials, dense gases, and far from low-temperatures ($20$ K). In the engineering practice, these constitutive equations are well-known and widely used. Nevertheless, there are situations where some extensions of them must be applied. Such a case could occur on small (micro or nano) length scales, short time scales, near low-temperature or far from equilibrium. 

Considering only heat conduction, various forms of heat propagation are experimentally found \cite{Ches63, GK66, JacWalMcN70, Rog71a, NarDyn72a, Gyar77a, JosPre89, JosPre90a, DreStr93a}. These are called second sound and ballistic propagation \cite{McNEta70a, McN74t, DreStr93a, KovVan15}. Their modeling background is diverse, and one can find many extended heat conduction equations with many different interpretations in the literature \cite{Van01a, VanFul12, Chen01, Cimm09diff, Van15a, JouEtal10b, SellEtal16b}. One of them is related to the approach of Rational Extended Thermodynamics (RET) \cite{DreStr93a, MulRug98}, it considers kinetic theory rigorously and uses phonon hydrodynamics to describe these deviations from Fourier's law \cite{KovVan16, KovVan18}. Another approach uses non-equilibrium thermodynamics with internal variables and current multipliers (NET-IV) \cite{KovVan15}. Both are tested on the same heat conduction experiment, and the latter one seems to be more effective \cite{KovVan18}. 

The main difference between these two approaches is routed to the physics laying behind the system of constitutive equations. Using the kinetic theory, one always has to assume apriori a mechanism occurring between the particles and describe the interaction among them. On the other side, non-equilibrium thermodynamics is phenomenological, the derivation of constitutive equations does not require any assumption regarding the microstructure, which makes the model more general and, in parallel, offers more degrees of freedom by not restricting the coupling coefficients. In the kinetic theory, due to the prescribed interaction model, most of the coefficients can be calculated, and only a few of them have to be fitted to the experimental data. Although its mathematical structure is advantageous, it is symmetric hyperbolic \cite{MulRug98, RugSug15}, and the fixed parameters lead to its weakness: e.g., in a previously mentioned heat conduction problem, one has to use at least $30$ momentum equations with increasing tensorial order to obtain the ballistic propagation speed approximately. The approach of NET-IV can resolve this problem, also preserving the structure of momentum equation; however, it requires more parameter to fit \cite{KovVan15, myphd2017, KovVan16, KovVan18}. All these approaches have advantages and disadvantages, and their detailed comparison is presented in \cite{KovEtal18rg}.

In case of investigating room temperature non-Fourier phenomenon, the phonon picture is not applicable \cite{myphd2017}. It is one advantage of NET-IV, it is applicable and tested on room temperature experiments that show over-diffusive type non-Fourier heat propagation \cite{Botetal16, Vanetal17, FulEtal18e}. It makes the kinetic approach of heat conduction more challenging to apply for practical problems; however, there are situations where its predictive power is useful (e.g., estimating transport coefficients). Such a situation is related to the topic of rarefied gases \cite{Struc05}. 
From some sense, the behavior of a rarefied gas (i.e., a gas under low pressure) is analogous with a rarefied phonon gas that applied in case of heat conduction. The difference among them is the type of the particle and the interpretation of some physical quantities. In order to understand the analogy, the ballistic conduction must be defined. 

Using phonon hydrodynamics, the ballistic heat conduction is interpreted as non-interacting particles that scatter on the boundary only, i.e., traveling through the material without any collision \cite{DreStr93a}. This assumption leads to the system of equations in one spatial dimension:
\begin{align}
\partial_t e + c^2 \partial_x p =& 0, \nonumber \\
\partial_t p + \frac{1}{3} \partial_x e +\partial_x N =& -\frac{1}{\tau_R} p, \label{eq_kt_ball} \\
\partial_t N + \frac{4}{15} c^2 \partial_x p =& - \left( \frac{1}{\tau_R}+\frac{1}{\tau_N} \right ) N, \nonumber
\end{align}
where $e$ being the energy density, $p$ is momentum density, $c$ stands for the Debye speed, $\tau_R$ and $\tau_N$ are the relaxation times referring to the resistive and normal processes \cite{DreStr93a}, furthermore, $\partial_t$ denotes the partial time derivative, applied for a rigid heat conductor. Here, $N$ is the deviatoric part of the pressure tensor. In phonon hydrodynamics, it can be identified as a current density of the heat flux. The key aspect to model ballistic effects is achieving coupling between the heat flux and the pressure. This is one merit of this approach: such coupling was not realized in any other theories before. That was the motivation for the approach of NET-IV, this coupling is obtained using current multipliers \cite{Nyiri91}, and the same structure can be reproduced \cite{KovVan15, myphd2017}:
\begin{align}
\rho c \partial_t T + \partial_x q = &0 \nonumber, \\
\tau_q \partial_t q +  q + \lambda \partial_x T + \kappa \partial_x Q =& 0  \nonumber , \\
\tau_Q \partial_t Q +  Q + \kappa \partial_x q=& 0 , \label{bc_eq3}
\end{align}
where $Q$ plays the role of $N$, $q$ is the heat flux, $c$ denotes the specific heat and the coefficient $\kappa$ is not fixed on contrary to (\ref{eq_kt_ball}), this property allows to adjust the exact propagation speed using only $3$ equations instead of $30$. The properties of the models above are discussed deeply by Jou et al. \cite{JouVasLeb88ext}, Alvarez et al. \cite{AlvEtal09} and Guo et al. \cite{GuoWang15f}.
The situation is the same for rarefied gases. Here, a gas under low pressure consists few enough particles to observe the ballistic propagation. In NET-IV, the starting point is the generalization of entropy density and its current:
\begin{align}
s(e, \rho, q_i, \Pi_{ij}) &=s_e(e, \rho) - \frac{m_1}{2} q_i q_i -\frac{m_2}{2} \Pi_{\langle ij \rangle} \Pi_{\langle ij \rangle} - \frac{m_3}{6} \Pi_{ii} \Pi_{jj}, \nonumber \\
J_i &= (b_{\langle i j \rangle} + b_{kk}\delta_{ij}/3)q_j,
\label{JC}
\end{align}
then exploiting the entropy production inequality of second law \cite{KovVan15}, one obtains a continuum model compatible with the kinetic theory to model rarefied gases \cite{myphd2017, KovEtal18rg}. Here $\Pi_{ij}$ is an internal variable \cite{BerVan15, BerVan17b, Verhas97}, it is identified as the viscous pressure, $\Pi_{ij}=P_{ij} - p\delta_{ij}$ with $p$ being the hydrostatic pressure, in accordance with EIT \cite{JouEtal10b, SellEtal16b}. This is the usual assumption in theories of Extended Thermodynamics, as a consequence of the compatibility with kinetic theory \cite{MulRug98, JouVasLeb88ext}, it also includes Meixner's theory \cite{Meix43a}, the first extension of Navier-Stokes equation. Besides, $b_{ij}$ is called Nyíri-multiplier (or current multiplier) \cite{Nyiri91} which permits obtaining coupling between the heat flux and the pressure. Furthermore, the form of entropy flux (\ref{JC}) is compatible with the one proposed by Sellitto et al. \cite{SellEtal13}, there the gradient of heat flux acts as a multiplier.  Since the proper description requires the separation of deviatoric and spherical parts, in eq.~(\ref{JC}) ${\langle \rangle}$ denotes the traceless part of the pressure. Einstein's summation convention is applied, too.
Equation (\ref{JC}) presents the same generalization as used for modeling ballistic heat conduction, thus, hereinafter it is called as \emph{ballistic generalization} of entropy and its current density \cite{myphd2017}. The linearized-generalized Navier-Stokes-Fourier system reads in one dimension \cite{myphd2017}:
\begin{align}
\tau_q\partial_t q +q +\lambda \partial_x T - \alpha_{21} \partial_x \Pi_s - \beta_{21} \partial_x \Pi_d =& 0, \nonumber \\
\tau_d \partial_t \Pi_d +\Pi_d + \nu \partial_x v + \beta_{12} \partial_x q =& 0, \nonumber \\
\tau_s \partial_t \Pi_s +\Pi_s + \eta \partial_x v + \alpha_{12} \partial_x q =& 0,
\label{NETSYSLINFINAL}
\end{align}
where the lower indices $d$ and $s$ denote the deviatoric and spherical parts, respectively. The $\eta$ is the bulk viscosity, $\nu$ denotes the shear viscosity, $\alpha_{ab}$, $\beta_{ab}$ ($a,b=1,2$) are the coupling parameters between the heat flux and the pressure and $\tau_m$ ($m=q,d,s$) are the relaxation times, here the coupling parameters and the relaxation times are to be fit. This structure is equivalent with the 1D linearized version model from RET \cite{Arietal12, AriEtal12c, Arietal13}:
\begin{align}
\tau_q \partial_t q + q + \lambda \partial_x T - R T_0 \tau_q \partial_x \Pi_d + R T_0 \tau_q \partial_x \Pi_s= &0, \nonumber \\
\tau_d \partial_t \Pi_d + \Pi_d + 2 \nu \partial_x v - \frac{2 \tau_d}{1+c_v^*}\partial_x q=&0, \nonumber \\
\tau_{s} \partial_t \Pi_s + \Pi_s + \eta \partial_x v + \frac{\tau_s (2c_v^*-3)}{3c_v^*(1 + c_v^*)} \partial_x q =&0, \label{ET_LINSYS2}
\end{align}
with $R$ being the gas constant and $c_v^*$ denotes the dimensionless specific heat: $c_v^*=c_v / R$. As it is apparent, only the relaxation times are free parameters, all the other coefficients are fixed. 
It is interesting to note that the system (\ref{ET_LINSYS2}) is derived by considering a doubled hierarchy of balance equations \cite{RugSug15, PavEtal13}. The reason behind that fact is related to the more degrees of freedom within polyatomic gases \cite{KovEtal18rg}. It is also important to note that it is not the only way for the kinetic theory: Lebon and Cloot derived a possible generalization using gradient terms as new variables \cite{LebCloo89} to model nonlocal phenomena.
In order to obtain a complete (closed) system of equations, beside the constitutive equations above, one has to use the balance laws as well:
\begin{align}
\partial_t \rho + \rho_0 \partial_x v =&0, \nonumber \\
\rho_0 \partial_t v + \partial_x \Pi_d + \partial_x \Pi_s +R T_0 \partial_x \rho + R \rho_0 \partial_x T = &0, \nonumber \\
\rho_0 c \partial_t T + \partial_x q + R \rho_0 T_0 \partial_x v = &0, \label{balances}
\end{align}
i.e., the mass, momentum and energy balances, respectively. 

It is worth mentioning the earlier works of Lebon and Cloot \cite{LebCloo89} and Carrassi and Morro \cite{CarMorr72a, CarMorr72b} where a similar comparison is performed. In these papers, other experiments are analyzed that conducted by Meyer and Sessler \cite{MeySess57}, which slightly differ from the following one.

One essential remark about the rarefied gases is connected to the classical literature of kinetic theory \cite{Klimontovich95b}. According to Klimontovich, a density parameter $\varepsilon$ should be small for rarefied gases: $\varepsilon<<1$. Using the air properties at atmospheric pressure and room temperature, it turns out that $\varepsilon\approx 10^{-4}$ \cite{Klimontovich95b}. In other words, air at 10 atm is also rarefied or at least close to a rarefied state. 
 In contrary to the classical literature of kinetic theory, a gas at 1 atm is considered to be a dense state. Moreover, in the presented experimental data, the pressure goes below 2000 Pa, and that state is called a rarefied one.

\section{Experiments}

As in the case of heat conduction \cite{KovVan18}, the experimental results are considered as a benchmark problem in order to test the validity and feasibility of the corresponding generalized model. Here, the measurement performed by Rhodes \cite{Rhod46} is discussed in detail. There are many other data in the literature \cite{SetEtal55, Gre56, SluiEtal64, SluiEtal65}, but this one is going to be sufficient to present all the necessary conclusions and difficulties arising in that field. 

\begin{figure}[H]
\includegraphics[width=8cm,height=5.7cm]{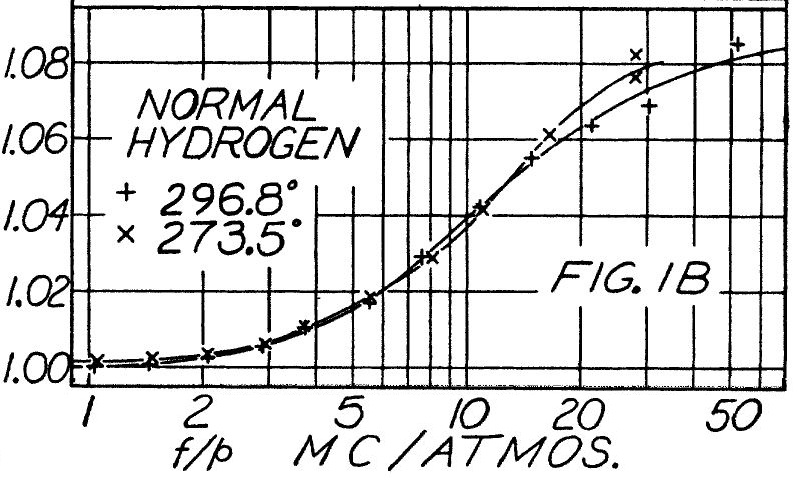}
\caption{Speed of sound measurement performed by Rhodes \cite{Rhod46}. The vertical axis denotes the relative speed of sound, i.e., $v/v_0$, where $v_0$ is the speed of sound related to the normal state.}
\label{rhodmeas}
\end{figure}

A sonic interferometer \cite{Stew46} is used to measure the sound speed for various frequency-pressure ratios \cite{Rhod46}, see Fig.~\ref{rhodmeas} for typical data. The interferometer is placed in a dewar to maintain a constant temperature within. 

It has to be emphasized that the frequency was constant as well in the experiments \cite{Rhod46}, i.e., the pressure is varied over the whole interval. More appropriately, it was the density that changed during the experiment when the constant temperature has maintained. In Fig.~\ref{rhodmeas}, the results related to normal Hydrogen is presented. The measurement is also performed using pure para-Hydrogen and the $50-50$ mixture of para-ortho Hydrogen \cite{Rhod46}. Now choosing the curve from Fig. \ref{rhodmeas} corresponding to $296.8$ K. Before making any advancement with the extended models, two essential aspects must be discussed. The first one is to investigate the density dependence of material parameters. Then, one can calculate the dispersion relation for the relating model (\ref{NETSYSLINFINAL}--\ref{balances} or \ref{ET_LINSYS2}-\ref{balances}) to model the experiment and analyze the frequency-pressure dependency, too. 

\subsection{Density dependence of material parameters}
Despite the fact that the experimental data is recorded as a function of $\omega/p$ only, it is also emphasized in the related paper of Rhodes \cite{Rhod46} that the frequency was constant all along the measurement in an isothermal environment. It means that the pressure is varied by changing the density of the gas only. That is, changing the state of the gas from normal (or dense) state to a rarefied one must reflect the role of density dependence of material parameters. 

Indeed, this is an efficient way to demonstrate the validity region of the classical Navier-Stokes-Fourier equations\footnote{As a side note: if the gas is already rarefied at 1 atm, then how could the classical Navier-Stokes-Fourier equations be applied? In the introduction of \cite{Struc05} their validity limit is said to be below Kn$\approx 0.05$.}. If one considers that all the material parameters are independent of the density, then it would suggest that the constitutive equations are applicable for any density. However, as the experiment shows, some effects become essential when the gas reaches its rarefied state. Density-dependent material coefficients can represent it: some of them make the related memory and nonlocal effects (couplings) negligible for dense states and worthwhile to account in rarefied states while the others change accordingly. 

\begin{figure}
\includegraphics[width=10cm,height=7cm]{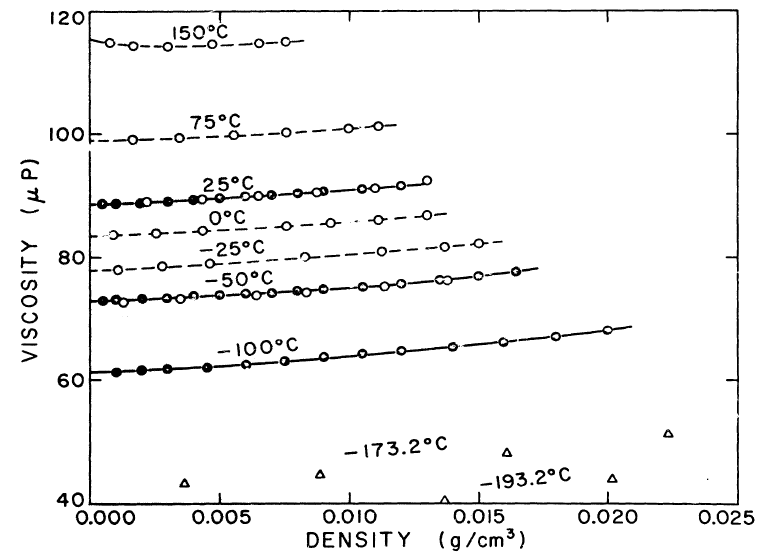}
\caption{Density dependence of viscosity for dense gases when the non-zero viscosity at zero density appears. \cite{GrackiEtal69a}.}
\label{grackimeas}
\end{figure}

Considering the RET approach of Arima et al. \cite{Arietal13}, one can notice the following:
\begin{itemize}
\item the viscosities are constant: $\nu=p_0 \tau_d$ for shear viscosity and $\eta \sim p_0 \tau_s$ for the bulk part,
\item the thermal conductivity is also constant: $\lambda \sim p_0 \tau_q$,
\item as a consequence of the above, the relaxation times are inversely proportional with mass density: $\tau \sim \frac{1}{\rho}$.
\end{itemize}
All the other parameters are fixed and do not depend on the mass density in any way. 
It is important to emphasize that the essential material parameters are constant for very different pressures, like $10^3$ or $10^5$ Pa. These are the consequences of the apriori assumptions; however, in general, it is not valid for the kinetic theory \cite{CohSan80, GulTrap86, GulTrap86a, GulSel02}. Constant, density-independent transport coefficients (viscosities and thermal conductivity) may be obtained for Maxwellian molecules \cite{TrueMunc80b}. In order to take into account their density dependence, the Enskog-type correction can be used \cite{Dym87, GulSel02, UmVes14}. It is in good agreement with particular experiments that are devoted to measuring the density dependence of shear viscosity for dense states \cite{GrackiEtal69a, GrackiEtal69b, Dym87}. However, it is important to mention here that these corrections are valid for dense gases and consider a constant term as a zero-density limit \cite{GrackiEtal69a, TrueMunc80b, Liboff03b}. Although the extrapolation from dense states (e.g., around few MPa) to rarefied states (e.g., around $10^2$-$10^3$ Pa and below) is not possible, it still reflects the presence of density dependence of transport coefficients (see Fig.~\ref{grackimeas} for details). 

On the other hand, the measurements of Itterbeek et al. \cite{ItterKee38, ItterCla38, IttPae40} demonstrate decreasing viscosity by decreasing the pressure. It can be only piecewise linear, its steepness changes drastically at very low pressures ($1-10$ Pa), and the viscosity tends to zero \cite{ittpaemeas}. This is contradictory to the zero density limit prediction from kinetic theory. 

\begin{figure}
\includegraphics[width=10cm,height=6cm]{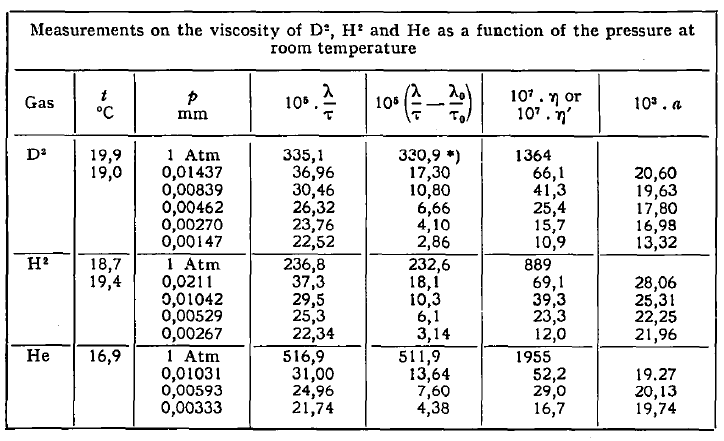}
\caption{Density dependence of viscosity for rarefied gases \cite{IttPae40}.}
\label{ittpaemeas}
\end{figure}

For the sake of complete comparison, we use the same assumptions, i.e.,
\begin{itemize}
\item both viscosities and the thermal conductivity are constant: \\ $\nu=8.82 \cdot 10^{-6}$ Pas, $\eta=326 \cdot 10^{-6}$ Pas and $\lambda=0.182$ W/(mK), respectively.
\item all relaxation times and the coupling coefficients have $1/\rho$ dependence.
\end{itemize}
These follow from a scaling requirement which discussed in the next section.
%In contrary to the evaluation using RET, both the viscosities and the thermal conductivity are assumed to be density-dependent in a different way:
%\begin{itemize}
%\item instead of dynamical viscosity, the kinematic viscosity is introduced as: $\nu=\rho \mu$ (and $\eta \sim \rho $) where $\mu$ is the kinematic and $\nu$ is the dynamical viscosity, i.e., it behaves like a linear dependence of $\rho$. At zero density, the dynamical viscosity tends to be zero. 
%\item the thermal conductivity depends inversely on $\rho$: $\lambda \sim \frac{1}{\rho}$, in order to model the ballistic effects appearing at low-pressure, i.e., the resistive collisions among the molecules are less frequent. 
%\item the relaxation times are inversely proportional with the mass density: $\tau \sim \frac{1}{\rho}$,
%\item in the coupling parameters the density dependence is taken into account according to the type of the quantitiy: viscosity-like coefficients depend on linearly and time-like coefficients depend on inversely. 
%\end{itemize}
%These differences result in a completely distinct approach between RET and NET-IV at the level of experiments and looses the compatibility with kinetic theory. Indeed, the framework of NET-IV assures this freedom by not assuming a-priori the exact collision mechanism, however, the resulting degrees of freedom (i.e., the number of free parameters) are much higher than in case of RET. 

\subsection{Frequency - pressure dependence}
It is stated frequently in the early experimental papers \cite{SetEtal55, Gre56, SluiEtal64} that the behavior of a gas depends on the ratio of frequency and pressure alone. Although it is seemingly correct based on the experimental data \cite{Rhod46}, it requires constant transport coefficients, $1/\rho$ dependence in the new parameters (relaxation times and coupling coefficients) and ideal gas state equation. These assumptions work for the corresponding evaluations; it could not be valid for an extensive pressure (or density, accordingly) domain, e.g., from $10$ Pa to $10^8$ Pa as the viscosities and the thermal conductivity do depend on the pressure. Moreover, it is not clear how that scaling would appear using a more general equation of state.

This scaling property can be easily demonstrated for the classical Navier-Stokes-Fourier model by calculating the dispersion relation using the previous assumptions.
%One can calculate the dispersion relation for Navier-Stokes-Fourier equations using assumptions from kinetic theory: constant material parameters and ideal gas for the equation of state, $p=\rho R T$. 
Then, there no one will find terms containing the frequency $\omega$ and the pressure $p$ separately, as follows. 

Assuming the common $e^{i ( \omega t - kx)}$ plane wave solution of the system (\ref{NFS}) with the usual wave number $k$ and frequency $\omega$,
\begin{align}
\partial_t \rho + \rho_0 \partial_x v =&0, \nonumber \\
\rho_0 \partial_t v + \partial_x \Pi_d + \partial_x \Pi_s +R T_0 \partial_x \rho + R \rho_0 \partial_x T = &0, \nonumber \\
\rho_0 c_V \partial_t T + \partial_x q + R \rho_0 T_0 \partial_x v = &0, \nonumber \\
q +\lambda \partial_x T =& 0, \nonumber \\
\Pi_d + \nu \partial_x v =& 0, \nonumber \\
\Pi_s + \eta \partial_x v =& 0,  \label{NFS}
\end{align}
and omitting the detailed derivation, one obtains the following expression for phase velocity $v_{ph} = \frac{\omega}{k}$:
\small
\begin{align}
&v_{ph}^2= \frac{c R T \rho ^2+R^2 T \rho ^2+i c \eta  \rho  \omega +i \lambda  \rho  \omega +i c \nu  \rho  \omega}{2 c \rho ^2} + \nonumber \\ &+i \frac{\sqrt{\rho ^2 \left(-4 c \lambda  \omega  (-i R T \rho +(\eta +\nu ) \omega )+\left(i R^2 T \rho -\lambda  \omega +c (i R T \rho -(\eta +\nu ) \omega )\right)^2\right)}}{2 c \rho ^2}. \label{vph}
\end{align}
\normalsize
Expanding all the terms within eq.~(\ref{vph}), the $v_{ph}=v_{ph}(\omega/p)$ dependence becomes visible and all the experimental data can be evaluated without calculating the pressure (or the mass density) independently from the frequency. 
In the case of the generalized NSF model, the situation is the same, and the previous assumptions ensure such scaling. 
Here, the final remark is made from an experimental point of view: the frequency and the pressure are separately controlled and should be documented in this way. Then, the pressure dependence in any parameter could be implemented without any problem, and the model would be free from ``unnatural'' assumptions that may be made unconsciously. Moreover, it could extend the validity region of this modeling approach.

%However, it does not inevitably hold for the extended constitutive equations (\ref{NETSYSLINFINAL}). There, following the same steps to calculate dispersion relations, terms like $R^2 T / \big (2 c (-i+\tau_q\omega ) (-i+\tau_s \omega ) (-i+\tau_d \omega ) \big)$ appear without pressure or mass density. It is evident that such expressions cannot produce pure $\omega/p$ dependence for constant material parameters. This structure itself is not extraordinary or problematic but it is when one does not know the exact value of frequency or pressure. This is one of the crucial parts in the evaluation of rarefied gas experiments presented in \cite{SluiEtal64}. 

\subsection{Evaluations, comparisons and conclusions}
\begin{figure}[H]
\includegraphics[width=10cm,height=7cm]{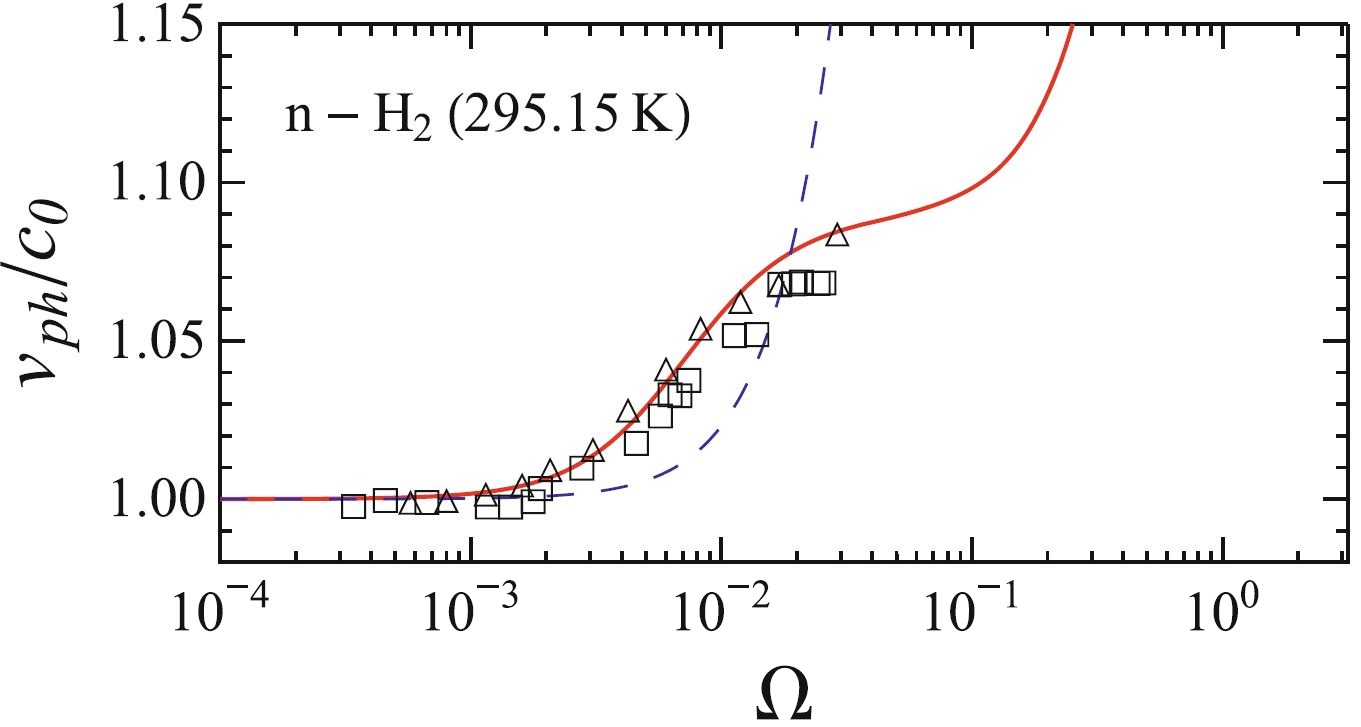}
\caption{Calculations of Arima et al. \cite{RugSug15}. The solid red line shows the prediction, the squares and triangles are referring to different experimental data; here, the triangles represent the data from Rhodes \cite{Rhod46}. The dashed line shows the behavior of the Navier-Stokes-Fourier equations.}
\label{m1}
\end{figure}

First, let us consider the results of Arima et al. \cite{Arietal13, RugSug15}, see Fig.~\ref{m1} for details. There are some important remarks on their evaluation method: 
\begin{enumerate}
\item Dimensionless frequency is used instead of $\omega/p$. Moreover, the papers \cite{Arietal13, RugSug15} do not reflect the fact that the frequency was constant, which could be confusing.
\item Regarding the temperature, $295.15$ K is used instead of $296.8$ K. It is seemingly a small difference; however, when one attempts to calculate the mass density from the data $\omega/p$, it leads to a different value, i.e., slightly inaccurate the data in \cite{RugSug15}. As a consequence, even the speed of sound is slightly inaccurate that used for vertical scale. %This is the reason why Fig.~\ref{m3} shows higher speed of sound than the measured one.
\item In \cite{Arietal13, RugSug15}, only certain ratios of relaxation times are fitted. However, no data is provided on how to calculate their exact value, and seemingly, the pressure is treated arbitrarily while it is directly given from the experimental data. 
\item One single parameter set is used to fit the data from papers \cite{Rhod46, WinHill67, SluiEtal64}. 
\end{enumerate}
Notwithstanding, Fig.~\ref{m1} demonstrates that kinetic theory can model the behavior of the gas in the rarefied state.

\begin{figure}[H]
\includegraphics[width=10cm,height=9cm]{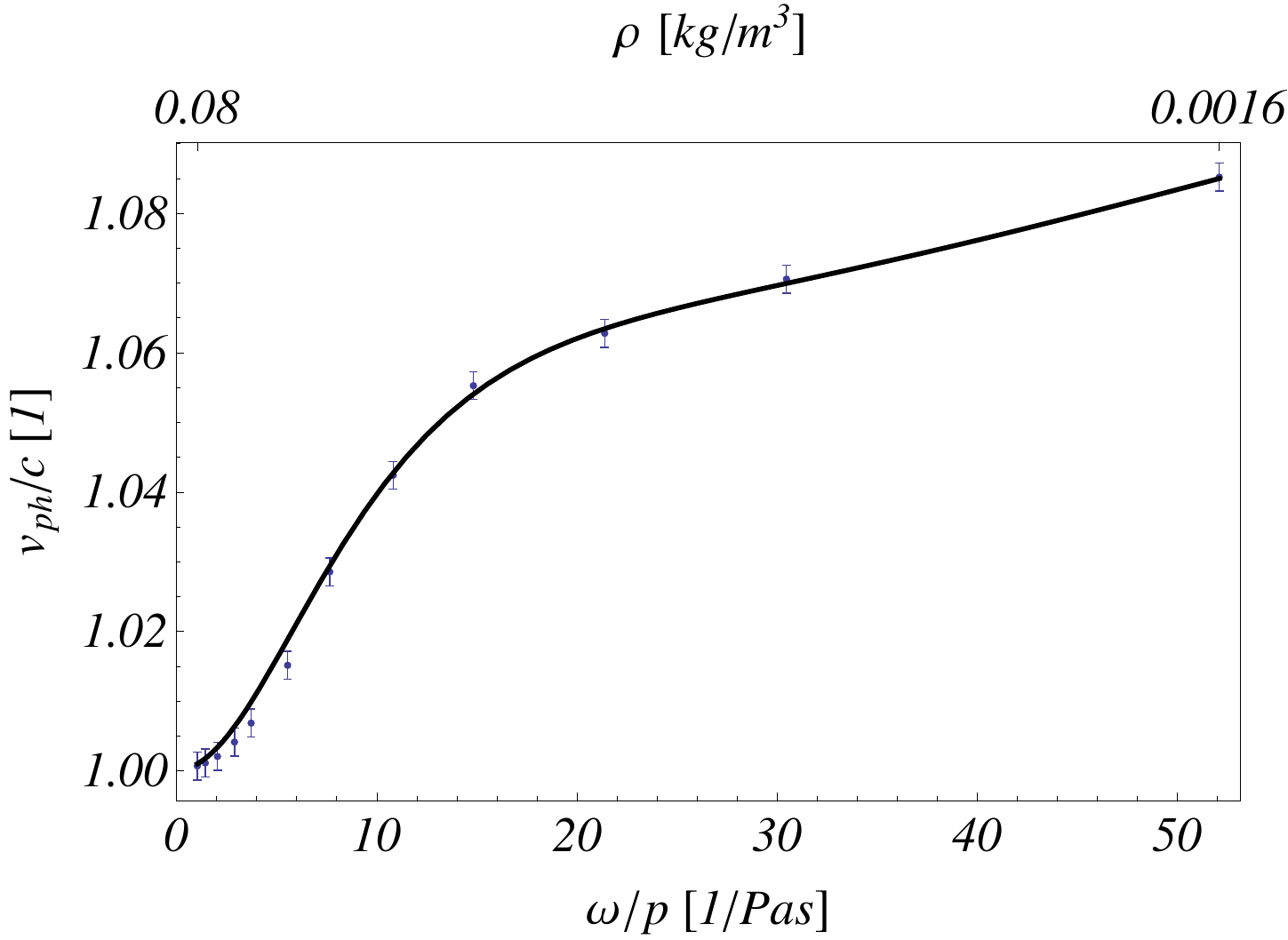}
\caption{Evaluation using NET-IV. The pressure starts at $1$ atm and decreases to $2000$ Pa, $\omega= 1$ MHz. Error bars are placed for each measurement point to indicate the uncertainty of digitalizing data, its magnitude is $\pm 2.5$ m/s.}
\label{m3}
\end{figure}

Here, using the NET-IV continuum model, only the experiment related to $T=296.8$ K is considered. In Figure \ref{m3}, two horizontal scales are used that intend to indicate the one-to-one correspondence between the $\omega/p$ and $\rho$. It is always possible if the frequency is known. Although the fitting procedure is conducted by hand, it is clear that the NET-IV model is also applicable to these problems. However, it is more difficult to do due to more degrees of freedom. Tables \ref{expcoeff1} and \ref{expcoeff2} show the corresponding values of each parameter.

\begin{table}
\caption{Fitted relaxation time coefficients for continuum model.} \label{expcoeff1}
\centering
\begin{tabular}{ccc} $\tau_q = \frac{t_1}{\rho}$, $t_1 = \left [\frac{kg \cdot s}{m^3} \right ]$ & $\tau_d = \frac{t_2}{\rho}$,  $t_2 = \left[\frac{kg \cdot s}{m^3}\right]$ & $\tau_s = \frac{t_3}{\rho}$ $t_3 =\left [\frac{kg \cdot s}{m^3}\right]$\\ \hline
$4 \cdot 10^{-8}$ &	$3.93 \cdot 10^{-8}$	& $8.8 \cdot 10^{-9}$\\
\end{tabular}
\end{table}

\begin{table}
\caption{Fitted coupling coefficients for continuum model.} \label{expcoeff2}
%\centering
\begin{tabular}{cccc} $\alpha_{12}= \frac{a_{12}}{\rho}$, $a_{12} =\left [\frac{kg \cdot s}{m^3}\right]$ &  $\beta_{12}= \frac{b_{12}}{\rho}$, $b_{12} =\left [\frac{kg \cdot s}{m^3}\right]$ & $\alpha_{21}=\frac{a_{21}}{ \rho}$, $a_{21} =\left [\frac{kg}{m \cdot s}\right]$ & $\beta_{21}=\frac{b_{21}}{ \rho}$, $b_{21} =\left [\frac{kg}{m \cdot s}\right]$ \\ \hline
$2.69 \cdot 10^{-5}$ &	$1.174 \cdot 10^{-4}$	& $8.323 \cdot 10^{-5} $ & $4 \cdot 10^{-6}$\\
\end{tabular}
\end{table}

%\begin{table}[H]
%\caption{Classical material parameters for continuum model.} \label{expcoeff3}
%\centering
%\begin{tabular}{ccc} $\nu=\mu \rho$, $\mu = \left [\frac{m^2}{s} \right ]$ & $\eta = \rho h$,  $h = \left[\frac{m^2}{s}\right]$ & $\lambda = \frac{l}{\rho}$ $l =\left [Pa^2 s K\right]$\\ \hline
%$ 10^{-4}$ &	$0.0036$	& $0.0175$\\
%\end{tabular}
%\end{table}

\section{Summary}
The models originated from RET and NET-IV are briefly introduced and tested on a particular experiment performed by Rhodes. It is apparent from the experimental data that the rarefaction in a Hydrogen gas shows substantial deviation from its dense state.
Here, the speed of sound data can be evaluated without any obstacles, but this evaluation highlights some crucial aspects. First, one has to clarify how the material parameters depend on the mass density. It is one of the most significant, cornerstone part of measurements. Second, all the data should be given in the experiment; the frequency-pressure ratio could be insufficient, especially when the transport coefficients are not constant respect to the pressure. It does matter what is changed: the frequency, the pressure, or both. 
In that sense, it is confusing to use dimensionless frequency, especially when all the necessary data is given for the appropriate scale. Besides that fact, dimensionless quantities are still valid to apply, but their use makes a bit more difficult to interpret the experimental data.

Moreover, it is also shown that besides the frequency/pressure scaling of experimental data, it is not necessarily obtained in the dispersion relations, merely by restrictive assumptions of constant coefficients. However, it is worth to note that if every experimental data is given appropriately, then the $\omega/p$ dependence does not stand as a necessary (or satisfactory) theoretical requirement.

The difference between the presented approaches originates in the preassumptions. Using kinetic theory, a detailed interaction is assumed, and it restricts some degrees of freedom. 
 On the other hand, in NET-IV, it is not necessary to make any preassumption regarding the type of the fluid, while in kinetic theory it is fixed as a first step, and the derived model such as eq.~(\ref{ET_LINSYS2}) will be valid only for this restricted case, in the price of the number of free coefficients.
As mentioned in the introduction, one of the main difficulty is the fitting of free parameters in the model of NET-IV (\ref{NETSYSLINFINAL}), that is, $7$ coefficients are to be fitted instead of $3$ that presented by eq.~(\ref{ET_LINSYS2}). However, this disadvantage could be an advantage if the number of equations is still enough to describe a three-dimensional process on the contrary to the kinetic theory where the momentum expansion generates a large system of equations for a more general situation.

\section{acknowledgement}
The author is thankful to Henning Struchtrup and Péter Ván for valuable discussions.
The work was supported by the grants NKFIH  K116197, K124366 and KH130378.

%\bibliographystyle{unsrt}
%\bibliography{references}

\end{document}